\documentclass[a4paper,11pt]{article}
\usepackage{pos}
\usepackage{bm}
\usepackage{amsmath}
\usepackage{amsfonts}
\usepackage{amsthm}
\usepackage{physics}
\usepackage{xcolor}
\usepackage{booktabs, subfig}
\usepackage[font=small,labelfont=bf]{caption}
\usepackage{hyperref}
\usepackage[compat=1.1.0]{tikz-feynman}

\usepackage{lipsum}

\let\OLDthebibliography\thebibliography
\renewcommand\thebibliography[1]{
  \OLDthebibliography{#1}
  \setlength{\parskip}{1.4pt}
  \setlength{\itemsep}{0.6pt plus 0.5ex}
}

\title{Direct lattice calculation of inclusive hadronic decay rates of the $\tau$ lepton}

\author*[a]{A.~Evangelista}
\author[a]{R.~Frezzotti}
\author[b]{G.~Gagliardi} 
\author[c]{V.~Lubicz}
\author[b]{F.~Sanfilippo}
\author[b]{S.~Simula}
\author[a]{N.~Tantalo}

\affiliation[a]{Dipartimento di Fisica and INFN, Università di Roma “Tor Vergata",
Via della Ricerca Scientifica 1, I-00133 Rome, Italy}
\affiliation[b]{Istituto Nazionale di Fisica Nucleare, Sezione di Roma Tre, Via della Vasca Navale 84, I-00146 Rome, Italy}
\affiliation[c]{Dipartimento di Matematica e Fisica, Università di Roma Tre and INFN, Sezione di Roma Tre, Via della Vasca Navale 84, I-00146 Rome, Italy}
  
\emailAdd{antonio.evangelista@roma2.infn.it}

\abstract{
The inclusive hadronic decay--rates of the $\tau$ lepton are particularly interesting from the phenomenological point of view since they give access to the CKM matrix elements $V_{ud}$ and $V_{us}$. In this talk, we discuss how a recent method for the extraction of smeared spectral densities from Euclidean lattice correlators can be used to obtain a direct lattice determination of inclusive hadronic $\tau$ decay rates. We also present preliminary numerical results obtained by applying this method to correlators measured on two gauge ensembles produced by the ETMC with $N_f=2+1+1$ dynamical flavours at physical pion masses, lattice spacing $a\simeq 0.08$~fm and volumes $L\simeq 5.1$~fm and $L\simeq 7.6$~fm.
}

\FullConference{%
  The 39th International Symposium on Lattice Field Theory (Lattice2022),\\
  8-13 August, 2022 \\
  Bonn, Germany 
}

\begin{document}
\maketitle

\section{Introduction}
Inclusive hadronic $\tau$ decays are particularly interesting from the phenomenological viewpoint since they give access to the CKM matrix elements $V_{ud}$ and $V_{us}$. The determinations of $V_{us}$ from leptonic and semileptonic kaon decays~\cite{Aoki:2021kgd} are in fairly good agreement with the one of Ref.~\cite{Hudspith:2017vew} but, for many years, a puzzling tension with other determinations obtained from inclusive hadronic $\tau$ decays  has been observed and debated~\cite{Maltman:2019xeh,Aoki:2021kgd}. On the lattice, hadronic $\tau$ decays have been studied by using dispersion relations and by combining non-perturbative lattice inputs with perturbative and/or OPE calculations (see for example~\cite{RBC:2018uyk}). 

Here we present a method to perform a fully non-perturbative direct lattice calculation of the $\tau\mapsto X \nu_\tau$ decay rate. In our method, the decay rate is extracted from the two-point Euclidean correlators of the hadronic weak currents that mediate the decay. This is done by using the algorithm of Ref.~\cite{Hansen:2019idp} that allows to extract smeared spectral densities from Euclidean lattice correlators and, building on Refs.~\cite{Gambino:2020crt,Gambino:2022dvu}, by using as smearing kernels smoothed versions of the step-functions that define the physical phase-space integration domain. 

We also present preliminary numerical results obtained by applying this method to the relevant correlators measured on two gauge ensembles produced by the Extended Twisted--Mass Collaboration (ETMC) with $N_f=2+1+1$ dynamical flavours with physical pion mass. The two ensembles, corresponding to the cB211.072.64 (B64 in short) and cB211.072.96 (B96 in short) entries in TABLE~V of Ref.~\cite{Alexandrou:2022amy}, have the same lattice spacing, $a=0.07957(13)$~fm, and differ only for the physical volumes that are $L=5.09$~fm and $L=7.64$~fm respectively. 

\section{Reconstruction of the inclusive rate using the HLT method }

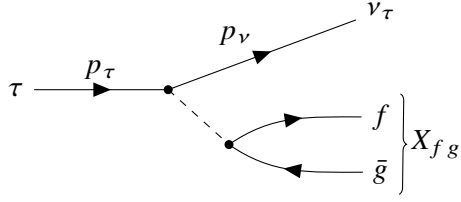
\begin{figure}
\centering
    \begin{tikzpicture}
        \begin{feynman}
            \vertex (a1) {\(\tau\)};
            \vertex [right=2.0cm of a1] (a2);
            \vertex [right=2.8cm of a2] (e1);
            \vertex [above=0.8cm of e1] (a3) {\(\nu_\tau\)};
            
            \vertex [below=0.056cm of e1] (a4) {\(f\)};
            \vertex [below=0.8cm of e1] (a5) {\(\bar{g}\)};
            \vertex at ($(a4)!0.5!(a5) - (2.0cm, 0)$) (e2);
        
            \diagram*{
                (a1) -- [fermion, edge label=\(p_\tau\)] (a2) [dot] ,
                (a2)  -- [fermion, edge label=\(p_\nu\)] (a3),
                (a2) [dot] -- [scalar] (e2) [dot],
                (a5) -- [fermion, out= 180, in=-35] (e2) [dot]-- [fermion, out=35, in=180] (a4),
            };
                        
            \draw [decoration={brace}, decorate] (a4.north east) -- (a5.south east)
                                                    node [pos=0.5, right] {\(X_{fg}\)};
            \draw [fill=black] (a2) circle(0.52mm);
            \draw [fill=black] (e2) circle(0.52mm);
        \end{feynman}
    \end{tikzpicture}
\caption{The $\tau\to\nu_\tau X_{fg}$ Feynman diagram (with no gluons). The hadronic final state $X_{fg}$, with flavour quantum numbers $f$ and $g$, has fixed 4-momentum $p_X = p_{\tau} - p_{\nu}$.}
\end{figure}

By relying on the Fermi effective theory for the weak interactions and by neglecting long--distance QED radiative corrections, the ratio $R_{fg}$ of the inclusive hadronic decay rate $\Gamma[\tau \mapsto X_{fg} \nu_\tau]$ with the leptonic decay rate $\Gamma[\tau \mapsto e\Bar{\nu}_e \nu_\tau]$ can be expressed as
\begin{flalign}
R_{fg}= 12\pi\,S_{EW}\abs{V_{fg}}^2\int_{r_{fg}}^1\dd\omega\,\omega\left(1 - \omega^2\right)^2\,\left\{\rho_{fg}^L(\omega) + \rho_{fg}^T(\omega)\left(1 + 2\omega^2\right)\right\}\;.
\end{flalign}
In the previous formula, $f$ and $g$ label the flavour quantum numbers of the final hadronic states $X_{fg}$ having four--momentum $p_X$, $r_{fg}=m_{fg}\slash m_\tau$ is the ratio of the mass of the lightest hadronic state and the $\tau$-mass, $S_{EW}= 1.0201(3)$ is the short--distance electroweak correction~\cite{Erler:2002mv}. 
The longitudinal and transverse form factors $\rho_{fg}^L(\omega)$ and $\rho_{fg}^T(\omega)$ parametrize the hadronic spectral density
\begin{flalign}
\mathcal{H}^{\mu\nu}_{fg}(p_X)
&=(2\pi)^4 \bra{0}H_{fg}^\mu(0)\, \delta^{(4)}(\mathbb{P}-p_X)\, H_{fg}^{\nu\dagger}(0)\ket{0}
\nonumber \\
&=p_X^\mu p_X^\nu\,\rho_{fg}^L(\omega) + \left[p_X^\mu p_X^\nu - g^{\mu\nu}p_X^2 \right]\rho_{fg}^T(\omega)\;,
\qquad
\omega^2=\frac{p_X^2}{m_\tau^2}\;,
\end{flalign}
where $\mathbb{P}=(\mathbb{H},\mathbb{\vec P})$ is the QCD four--momentum operator and $H_{fg}^\mu = V_{fg}^\mu-A_{fg}^\mu$ is the hadronic weak current that mediates the decay. 

In the following, we concentrate on the $ud$-flavour channel and omit the $fg$ flavour indexes in intermediate expressions. Moreover, we study separately the longitudinal ($L$) and transverse ($T$) contributions to $R_{ud}$ and also the contributions coming from the vector ($V^\mu$) and axial-vector ($A^\mu$) currents. To this end we introduce the indexes
\begin{flalign}
I=\{L,T\}\;,
\qquad
J=\{ V, A \}\;,
\end{flalign}
and the different components of the spectral density $\mathcal{H}^{\mu\nu}(p_X)$ according to
\begin{flalign}
\mathcal{H}^{L}_{J}(\omega)\equiv\mathcal{H}^{00}_{J}(\omega)\;,
\qquad
\mathcal{H}^{T}_{J}(\omega)\equiv\frac{1}{3}\sum_{i=1}^3\mathcal{H}^{ii}_{J}(\omega)\;,
\qquad
\mathcal{H}^{I}(\omega)\equiv
\mathcal{H}^{I}_V(\omega)+\mathcal{H}^{I}_A(\omega)\;,
\end{flalign}
with
\begin{flalign}
&
\mathcal{H}^{\mu\nu}_{J}(p_X)
\equiv (2\pi)^4 \bra{0}J^\mu(0)\, \delta^{(4)}(\mathbb{P}-p_X)\, J^{\nu\dagger}(0)\ket{0}~.
\end{flalign}
By working in the reference frame where the final hadronic state is at rest,
\begin{flalign}
p_X=(m_\tau\omega,\vec 0)\;,
\end{flalign}
we have
\begin{flalign}
&
R^I_J(\sigma)=12\pi\,S_{EW}\abs{V_{ud}}^2 \int_{r_{ud}}^\infty\dd\omega\,
\mathcal{H}^{I}_{J}(\omega)\,K_\sigma^I(\omega)~,
\nonumber \\
\nonumber \\
&
R_{ud}
= \lim_{\sigma\to 0} R_{ud}(\sigma)
= \lim_{\sigma\to 0}\left\{ 
R_V^L(\sigma)
+
R_A^L(\sigma)
+
R_V^T(\sigma)
+
R_A^T(\sigma)
\right\}
\;,
\label{eq:Rgfrep}
\end{flalign}
where, in analogy to Refs.~\cite{Gambino:2020crt,Gambino:2022dvu}, we have introduced the longitudinal ($K_{\sigma}^{L}$) and transverse ($K_{\sigma}^{T}$) smearing kernels
\begin{flalign}
K_\sigma^L(\omega)=\frac{(1-\omega^2)^2}{\omega}\Theta_\sigma(1-\omega)\;, \quad K_\sigma^T(\omega)=\frac{(1-\omega^2)^2(1+2\omega^2)}{\omega}\Theta_\sigma(1-\omega)\;.
\end{flalign}
The function $\Theta_{\sigma}(x)$ appearing in the previous formula can be any $C_\infty$ smoothed version of the step-function $\theta(x)$ such that $\lim_{\sigma\mapsto 0}\Theta_\sigma(x)=\theta(x)$.  In the following, we will consider the three different choices given by
\begin{flalign}
\label{eq:sm_kernels}
\Theta^{(1)}_\sigma(x)=\frac{1}{1+e^{-x\slash\sigma}} \;,
\quad
\Theta^{(2)}_\sigma(x)=\frac{1}{1+e^{-\sinh\left(x\slash\sigma\right)}} \;,
\quad
\Theta^{(3)}_\sigma(x)=\frac{1+\text{Erf}\left(x\slash\sigma\right)}{2} \;.
\end{flalign}
Under the assumption that the spectral densities $\mathcal{H}^I(\omega)$ are regular at the end-point of the phase-space, i.e. $\omega=1$, an analytical calculation shows that the corrections to the $\sigma\mapsto 0$ limit are even functions of $\sigma$, starting at $\order{\sigma^4}$, i.e.
\begin{flalign}
\int_{r_{ud}}^\infty\dd\omega\,\mathcal{H}^{I}(\omega)
\left\{
K_\sigma^{I}(\omega)-
K_0^{I}(\omega)
\right\}=\mathcal{O}(\sigma^4)\;,
\label{eq:volumescaling}
\end{flalign}
This assumption is of course not valid on a finite volume where the spectral densities are not regular. Indeed, because of the quantization of the spectrum, the finite--volume spectral densities $\mathcal{H}^{I}(\omega)$ are sums of Dirac $\delta$-functions localized in correspondence of the eigenvalues of the finite--volume Hamiltonian. However, precisely for this reason and as emphasized in Ref.~\cite{Hansen:2019idp}, the $\sigma\to 0$ limit in Eq.~(\ref{eq:Rgfrep}) has to be taken \emph{after} performing the necessary $L\to\infty$ extrapolation of the lattice data. A detailed numerical investigation of the dependence upon the volume of our results is postponed to a future publication. Here, see below, we simply check that the results obtained on the two ensembles with volumes $L\simeq 5.1$~fm and $L\simeq 7.6$~fm are compatible within the statistical uncertainties and then attempt a $\sigma\to 0$ extrapolation by relying on Eq.~(\ref{eq:volumescaling}).   

The representation of $R_{ud}$ given in Eq.~(\ref{eq:Rgfrep}) allows for a straightforward application of the method developed in Ref.~\cite{Hansen:2019idp} along the lines of Ref.~\cite{Gambino:2022dvu}. The starting point is the relation between the hadronic spectral density  $\mathcal{H}^{\mu\nu}_{J}(p_X)$ and the Euclidean two-point correlator $C_{J}^{\mu\nu}$ at vanishing three-momentum (our lattice input), i.e.
\begin{flalign}
\label{eq:spec_dens_corr_rel}
C^{\mu\nu}_J(t)
\equiv \int\dd[3]{x} \mathrm{T}\bra{0}J^\mu(at,\Vec{x})J^{\nu\dagger}(0)\ket{0}
=\frac{m_\tau}{2\pi}\int_{r_{ud}}^\infty d\omega\, \, \mathcal{H}^{\mu\nu}_{J}(p_X)\, e^{-am_\tau \omega t},
\quad
p_X=(m_\tau\omega,\vec 0),
\end{flalign}
where $t$ is the Euclidean time in units of the lattice spacing $a$.\footnote{On a lattice having a finite temporal extent $T$, Eq.~(\ref{eq:spec_dens_corr_rel}) must be modified replacing in the r.h.s. $e^{-am_{\tau}\omega t}$ with $e^{-am_{\tau}\omega t}+ e^{-am_{\tau}\omega (T-t)}$. }  The main idea is then to express the smeared-kernels $K_{\sigma}^{L}(\omega)$ and $K_{\sigma}^{T}(\omega)$ in terms of the basis function $\{ e^{-am_{\tau}\omega t}\}_{t=1,\ldots, \infty}$, i.e.
\begin{flalign}
\label{eq:basis_expansion}
K_{\sigma}^{I}(\omega) = \sum_{t=1}^{\infty} g^{I}(t,\sigma) e^{-am_{\tau}\omega t}~.
\end{flalign}
In this way, once the coefficients $g^{I}(t,\sigma)$ are known, the longitudinal ($R^{L}_J$) and transverse ($R^{T}_J$) contributions to $R_{ud}$ can be computed from the knowledge of 
\begin{flalign}
C_{J}^{L}(t)=-\frac{2\pi}{m_\tau} C_{J}^{00}(t)\;,
\qquad
C_{J}^{T}(t)=\frac{2\pi}{3m_\tau} \sum_{i=1}^3C_{J}^{ii}(t)\;,
\end{flalign}
by using
\begin{flalign}
\sum_{t=1}^{\infty}g^{I}(t,\sigma)C^{I}_{J}(t)
=
\int_{r_{ud}}^\infty\dd\omega\,\mathcal{H}^{I}_{J}(\omega)\,K_\sigma^{I}(\omega)\;,
\label{eq:K_sum}
\end{flalign}
and inserting the result in Eqs.~(\ref{eq:Rgfrep}).
However, as discussed thoroughly in Refs.~\cite{Hansen:2019idp}, the problem of finding the coefficients $g^{I}(t,\sigma)$ presents a certain number of technical difficulties. First of all, the sums appearing on the r.h.s. of Eqs.~(\ref{eq:basis_expansion}) need necessarily to be truncated at a finite value $t=t_{max}$, hence the goal is to find a finite set of coefficients $g^{I}(t, \sigma)$, with $t \in \{ 1, \ldots,  t_{max}\}$, such that both the statistical (due to the fluctuation of $C_{J}^{I}(t)$) and the systematic errors (due to the inexact reconstruction of the kernels) in the resulting determination of $R_{J}^{I}$ are under control. If we were only concerned with systematic errors, the best coefficients $g^{I}(t,\sigma)$ could be obtained by minimizing the 
 quadratic form
\begin{flalign}
\label{eq:func_A}
A_{\alpha}^I[\boldsymbol{g}]=
\int_{E_{0}}^\infty\dd\omega\, e^{am_\tau\omega \alpha} \abs{ f(\omega; \boldsymbol{g}) - K^{I}_\sigma(\omega)}^2~,
\end{flalign}
with 
\begin{flalign}
f(\omega; \boldsymbol{g}) \equiv \sum_{t=1}^{t_{\text{max}}}g(t,\sigma)e^{-am_\tau\omega t}\;.
\end{flalign}
Indeed, for any $\alpha<2$ and $0<E_{0}< r_{ud}$, the functional in Eq.~(\ref{eq:func_A}) corresponds to a weighted $L_2$-norm in the functional space defined in the interval $[E_0,\infty]$.  However, for small values of $\sigma$, the coefficients $g^{I}(t,\sigma)$ resulting from the minimization of $A_{\alpha}^{I}[\boldsymbol{g}]$ turn out to be very large in magnitude and oscillating in sign, strongly amplifying the statistical errors of $C^{I}_{J}(t)$ when the $t_{\text{max}}$-truncated version of the sum in Eq.~(\ref{eq:K_sum}) is evaluated (see Ref.~\cite{Hansen:2019idp} for more details on this point). 
 
The method of Ref.~\cite{Hansen:2019idp}, provides a regularization mechanism to this problem, enabling to find an optimal balance between statistical and systematic errors. This is achieved by minimizing a linear combination
\begin{flalign}
W_{\alpha}^{IJ}[\boldsymbol{g}] \equiv \frac{A_\alpha^I[\boldsymbol{g}]}{A_{\alpha}^I[\boldsymbol{0}]} + \lambda B^{IJ}[\boldsymbol{g}]\;,
\end{flalign}
of the norm-functional $A_{\alpha}^{I}[\boldsymbol{g}]$ and of the error-functional
\begin{flalign}
B^{IJ}[\boldsymbol{g}] = \frac{1}{(C_{J}^{I}(0))^{2}} \sum_{t_1 , t_2 = 1}^{t_{max}} g(t_{1},\sigma)\, g(t_{2},\sigma)~ {\rm{Cov}^{I}_{J}}(t_1 , t_2 )~,
\end{flalign}
where ${\rm{Cov}^{I}_{J}}(t_1, t_2 )$ is the covariance matrix of the lattice correlator $C_{I}^{J}(t)$, and $\lambda$ is the so-called trade-off parameter~\cite{Hansen:2019idp}. For any fixed value of the algorithmic parameters $\boldsymbol{p} \equiv \{ \alpha, E_{0}, \lambda, t_{max}\}$, the minimization 
\begin{flalign}
\left. \frac{\partial W_{\alpha}^{IJ}[\boldsymbol{g}]}{\partial g(t,\sigma)}\right\vert_{\boldsymbol{g}=\boldsymbol{g}_{\boldsymbol{p}}^{IJ}}=0\;,
\end{flalign}
defines the coefficients $\boldsymbol{g}^{IJ}_{\boldsymbol{p}}$. The systematic error associated to the inexact reconstruction of the smeared kernel, 
\begin{flalign}
K^{IJ}_{\boldsymbol{p}}(\omega) \equiv f(\omega; \boldsymbol{g}^{IJ}_{\boldsymbol{p}}) = \sum_{t=1}^{t_{\text{max}}}g^{IJ}_{\boldsymbol{p}}(t,\sigma)e^{-am_\tau\omega t}\;,
\end{flalign}
can be quantified through the quantity
\begin{flalign}
\label{eq:dg_def}
d^{I}(\boldsymbol{g}) = \sqrt{ \frac{A_{0}^I[\boldsymbol{g}]}{A_{0}^I[\boldsymbol{0}]}}\;.
\end{flalign}
In the following, we will quote our best estimate for the four contributions $R_{V,A}^{L,T}(\sigma)$, see Eq.~(\ref{eq:Rgfrep}), performing the so-called \textit{stability analysis} (see Ref.~\cite{Bulava:2021fre} and also the Supplementary Material of Ref.~\cite{Alexandrou:2022tyn}), which amounts to select the algorithmic parameters $\boldsymbol{p}$ in such a way that the corresponding $d^I(\boldsymbol{g}^{IJ}_{\boldsymbol{p}})$ is sufficiently small and the results stable, within statistical errors, under variations of $\boldsymbol{p}$ (the so-called statistically dominated regime).\footnote{More numerical details on this point will be given in a forthcoming publication.}

\section{Numerical results}
\begin{figure}
\centering
\subfloat{\includegraphics[width=1.00\linewidth]{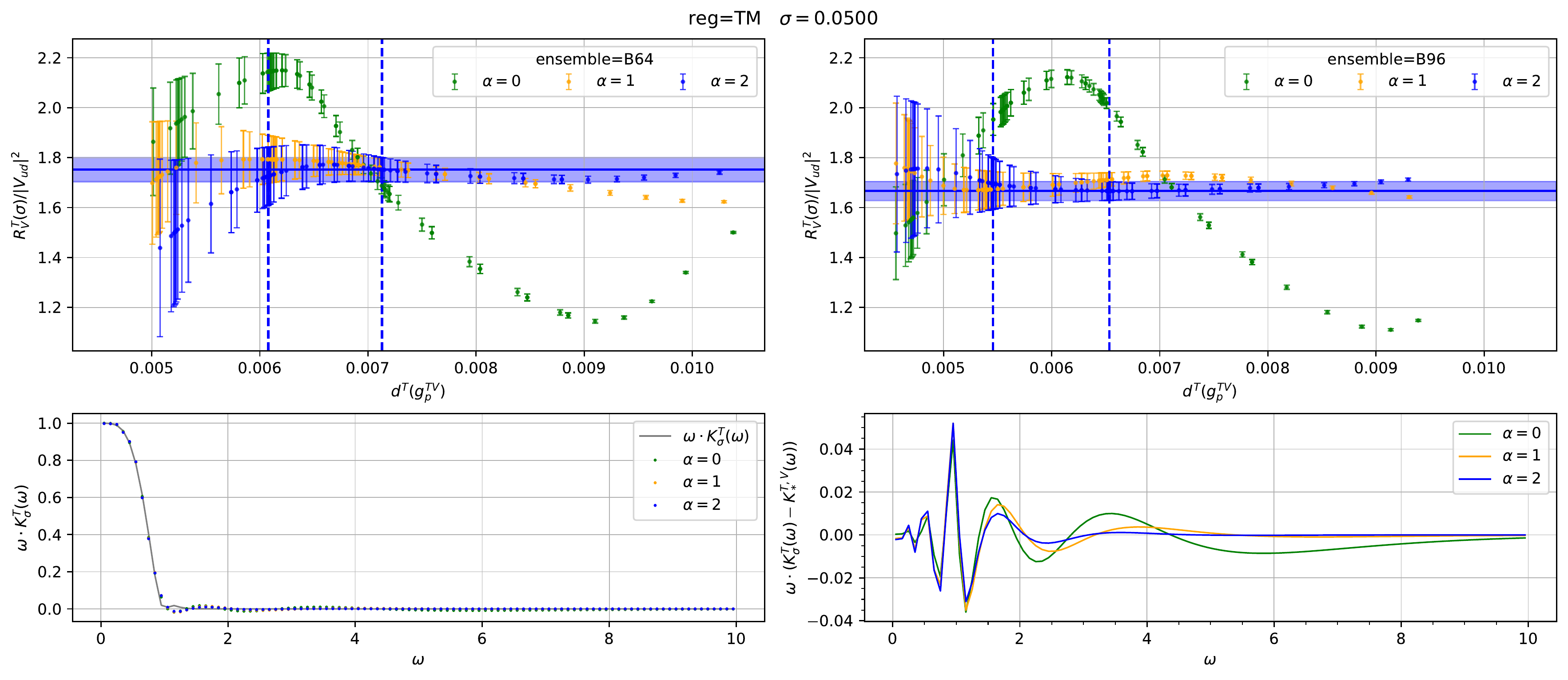}}
\caption{ {\it Top:} the contribution $R^{T}_{V}\slash\abs{V_{ud}}^2$  obtained using $\alpha=0$ (green), $\alpha=1$ (yellow) and $\alpha=2^-$ (blue), is plotted against $d^T(\boldsymbol{g}_{\boldsymbol{p}}^{TV})$ for $\sigma=0.05$. For $\alpha=2^{-},$ the rightmost (leftmost) vertical dashed line indicates the point satisfying Eq.~(\ref{Eq:res_condition}) with $r=10^4~(10^3)$, while the horizontal blue band corresponds to our final determination obtained combining in quadrature the statistical and the systematic errors. The results are shown in the TM lattice regularization for both the B64 (top-left figure) and the B96 (top-right figure) ensembles at $\sigma=0.05$. {\it Bottom:} the reconstructed smearing kernels $K^{TV}_{\boldsymbol{*}}(\omega)$, obtained using the coefficients $\bm{g_{*}}^{TV}$ of Eq.~(\ref{Eq:res_condition}) are compared, for $\alpha=0,1,2^-$, with the target one $K_{\sigma}^{T}(\omega)$ for $\sigma=0.05$ (bottom-left figure). In the bottom-right figure we show  $\omega\cdot( K_{\sigma}^{T}(\omega) - K_{\boldsymbol{*}}^{TV}(\omega))$.    }
\label{Fig:stability}
\end{figure}
\begin{figure}
\centering
\subfloat{\includegraphics[width=0.7\linewidth]{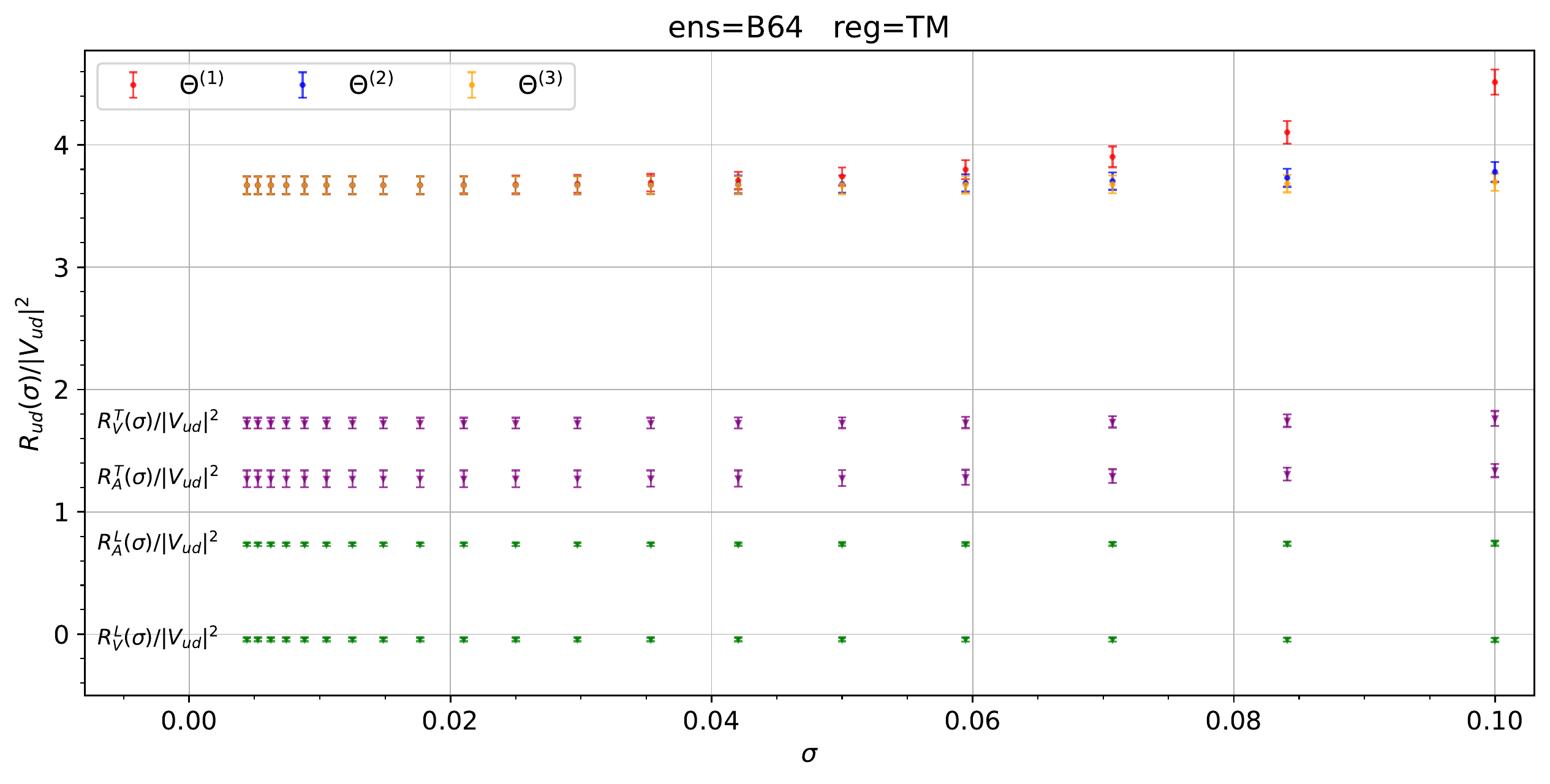}} \\ 
\vspace{-0.4cm}
\subfloat{\includegraphics[width=0.7\linewidth]{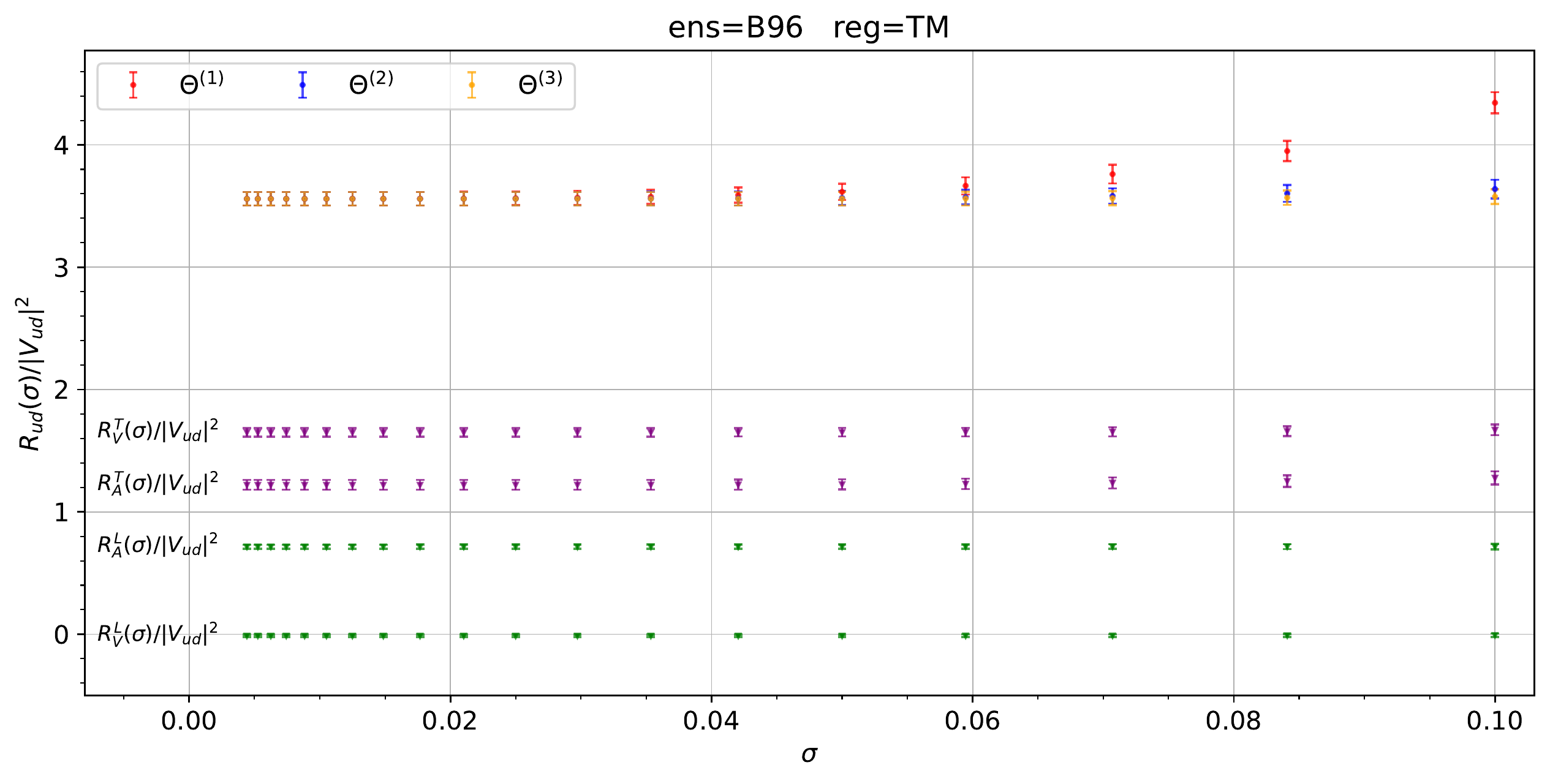}}
\caption{
The decay rate $R_{ud}(\sigma)\slash\abs{V_{ud}}^2$ as a function of $\sigma$ in the range $[0.0044,0.1]$. The results have been obtained in the TM regularization and are shown for both the volumes (B64 top, B96 bottom) and for the three choices of $\Theta(\omega)$ in Eqs.~(\ref{eq:sm_kernels}). In the case $\Theta^{(1)}_\sigma(\omega)$ we also show, separately, the four contributions  $R_{V,A}^{L,T}(\sigma)\slash\abs{V_{ud}}^2$.
}
\label{Fig:Results}
\end{figure}
\begin{figure}
\centering
\subfloat{\includegraphics[width=0.7\linewidth]{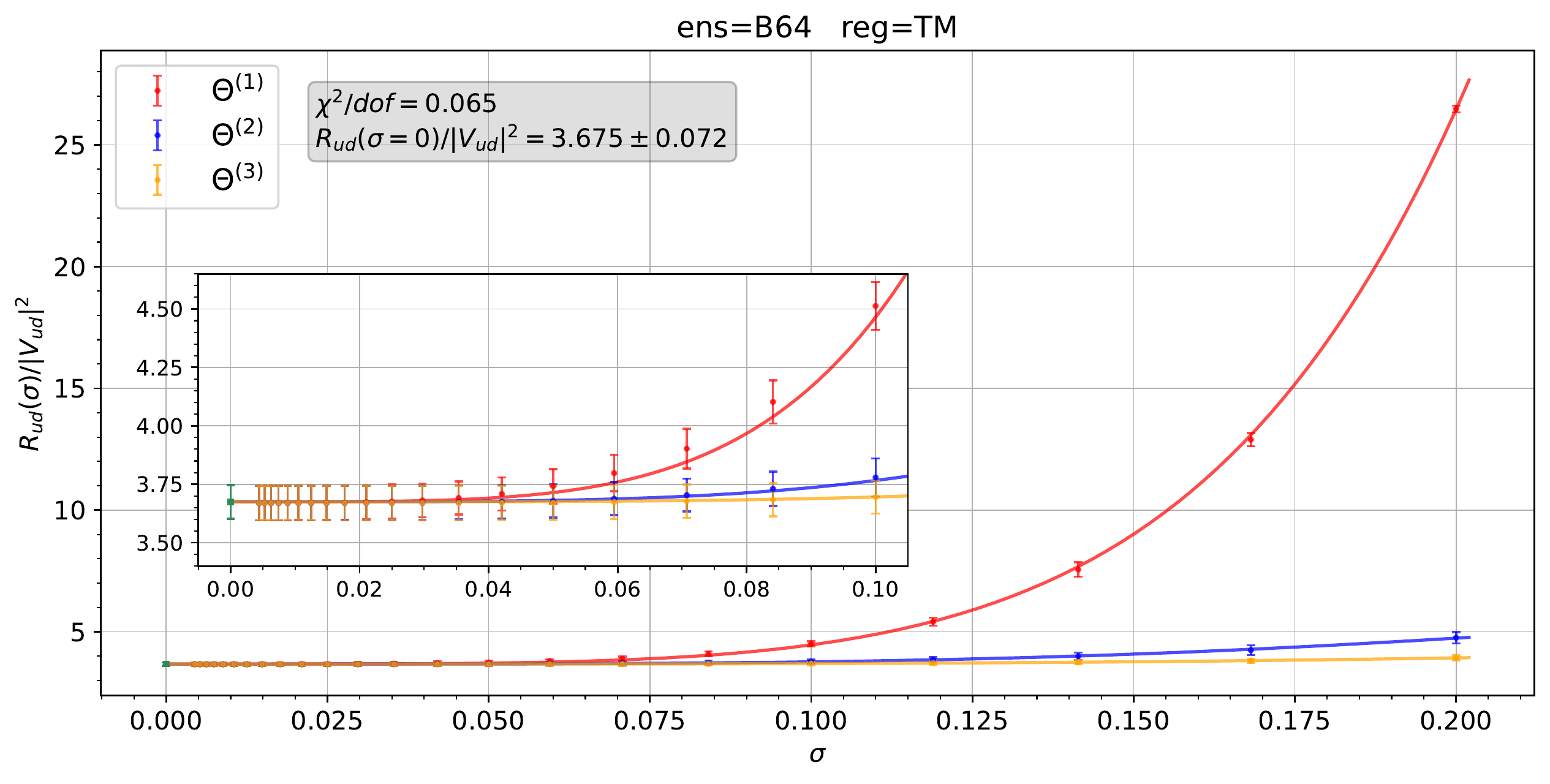}}\\
\vspace{-0.4cm}
\subfloat{\includegraphics[width=0.7\linewidth]{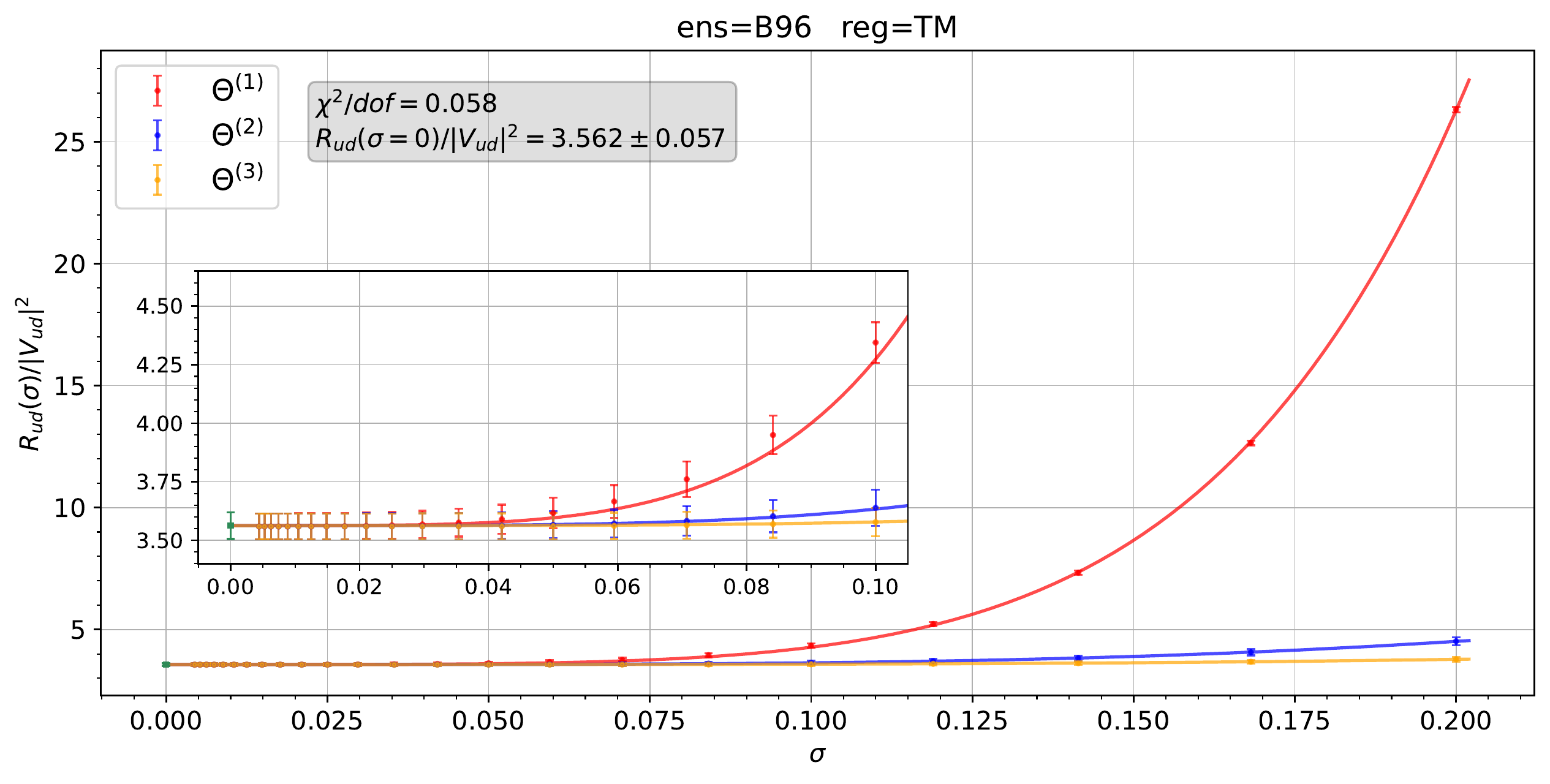}}
\caption{
Combined $\sigma\to 0$ extrapolations of our results for $R_{ud}(\sigma)\slash\abs{V_{ud}}^2$ obtained in the TM regularization for both volumes. The datasets corresponding to the three choices of $\Theta(\omega)$ appearing in Eqs.~(\ref{eq:sm_kernels}) have different colours. Assuming negligible finite-volume effects, these are expected to have the same $\sigma\to 0$ limit and to differ at finite $\sigma$ with leading corrections of $O(\sigma^4)$. The data have been fitted using the ansatz of Eq.~(\ref{eq:fit}). The green point is the result of the extrapolation while the solid curves are  the fitted curves $R_k(\sigma)$ for $k=1$ (red), $k=2$ (blue) and $k=3$ (yellow).
}
\label{fig:extrapolation}
\end{figure}
In this section, we present our preliminary results for $R_{ud}$. These have been obtained by using the Euclidean lattice correlators $C_{I}^{J}(t)$ produced by the ETMC on the two ensembles B64 and B96. We have considered two different discretized versions of the local weak current, peculiar to our twisted-mass LQCD setup, that in the following will be indicated as twisted-mass (TM) and Osterwalder-Seiler (OS)\,\cite{Frezzotti:2004wz}. The results obtained using the two discretizations only differ by $\mathcal{O}(a^{2})$ cut-off effects, enabling us to approach the continuum limit in two different ways. Furthermore, we considered three different values, 
\begin{flalign}
\label{eq:weights}
\alpha = \{ 0, 1 , 2^- \}\;,
\end{flalign}
for the parameter $\alpha$ appearing in Eq.~(\ref{eq:func_A}), where $\alpha=2^-$ in practice means $\alpha=1.99$. 
We set the parameter $E_{0}$ in Eq.~(\ref{eq:func_A}) to $E_{0}=0.05\simeq 0.6\, m_\pi/m_{\tau}$ and use $t_{max}=64,96$ respectively for the ensembles B64 and B96.

In Figure~\ref{Fig:stability} we show our determination of $R_{V}^{T}(\sigma)/|V_{ud}|^{2}$ in the TM regularization and at $\sigma=0.05$, obtained employing the three values of $\alpha$ and the smeared kernel $\Theta_{\sigma}^{(1)}$, see Eqs.~(\ref{eq:sm_kernels}).
The results are shown as a function of the parameter $d^T(\boldsymbol{g}_{\boldsymbol{p}}^{TV})$ defined in Eq.~(\ref{eq:dg_def}) and provide an illustrative example of our \textit{stability analysis}. For large values of $d^T(\boldsymbol{g}_{\boldsymbol{p}}^{TV})$ the results corresponding to different values of $\alpha$ are substantially different because in this regime the reconstruction of the smearing kernel is very bad. 
At very small values of $d^T(\boldsymbol{g}_{\boldsymbol{p}}^{TV})$, where the quality of the reconstruction becomes excellent, the results corresponding to the different values of $\alpha$ become compatible because the statistical errors are quite large. We observe that the results corresponding to $\alpha=1, 2^{-}$ stabilize at much larger values of $d^T(\boldsymbol{g}_{\boldsymbol{p}}^{TV})$ than the $\alpha=0$ ones. This behaviour, already observed in Ref.~\cite{Alexandrou:2022tyn} where the same $L_2$-norms have been used, can be explained by noticing that for $\alpha>0$ the presence of the exponential $e^{am_\tau\omega \alpha}$ in Eq.~(\ref{eq:func_A}) improves the quality of the reconstruction in the large-$\omega$ region. Indeed,  the errors in the reconstruction of the smearing kernels (e.g.\ $K_\sigma^T(\omega)$) for large values of $\omega$ get amplified in the corresponding smeared quantities (e.g.\ $R_J^T(\sigma)$) because, in general, spectral densities grow asymptotically with the energy (e.g.\ $\mathcal{H}^{T}_J(\omega)\propto \omega^{2}$).

For $\alpha=1,2^{-}$, we found that the results obtained at the point $\boldsymbol{g}_{\boldsymbol{*}}^{IJ}$ such that the condition
\begin{flalign}
\frac{A_\alpha^I[\bm{g_*}^{IJ}]}{A^I_\alpha[\bm{0}]} = r B^{IJ}[\bm{g_*}^{IJ}]~,
\qquad
r=10^4\;,
\label{Eq:res_condition}
\end{flalign}
holds true, are in the statistically dominated regime. In what follows, the central values of the four contributions to $R_{ud}/|V_{ud}|^{2}$ in Eq.~(\ref{eq:Rgfrep}) are estimated by using the $\alpha=2^{-}$ results (that are remarkably stable) and the coefficients $\boldsymbol{g}_{\boldsymbol{*}}^{IJ}$. Residual systematic errors are instead evaluated by re-performing the analysis using $r=10^{3}$ (see the vertical lines in Figure~\ref{Fig:stability}). Any variation of the result corresponding to the choice $r=10^{3}$ w.r.t. the result corresponding to $r=10^{4}$ that goes beyond a mere statistical fluctuation is added in quadrature to the statistical error.

In Figure~\ref{Fig:Results} we show our preliminary results for $R_{ud}(\sigma)/|V_{ud}|^{2}$ obtained in the TM regularization by using the three different smearing kernels of Eq.~(\ref{eq:sm_kernels}) and 23 values of $\sigma$ in the range 
\begin{flalign}
\sigma\in [0.0044,0.2]\;.
\end{flalign}
We observe a remarkably flat behaviour for $\sigma < 0.05$~\footnote{According to our experience, see e.g. Refs.~\cite{Bulava:2021fre,Gambino:2022dvu,Alexandrou:2022tyn}, the numerical reconstruction of smearing kernels corresponding to $\theta$-functions in the $\sigma\to 0$ limit is much easier than in the case of kernels corresponding to Dirac $\delta$-functions.}. Moreover, the results corresponding to the two volumes $L=5.1~{\rm fm}$ and $L=7.6~{\rm fm}$ are compatible at all values of $\sigma$ within less than 1.5 standard deviations. This implies that finite-volume effects are negligible within the quoted errors, even at the smallest value of $\sigma$ that we have considered. In the light of these observations, we attempted a combined $\sigma\to 0$ extrapolation of our results by relying on the infinite-volume asymptotic formula of Eq.~(\ref{eq:volumescaling}). On each ensemble and for each regularization, the results corresponding to the three smearing kernels $\Theta_{\sigma}^{(k)}$ ($k=1,2,3$) have been fitted by using the following ansatz
\begin{flalign}
R_k(\sigma)=R + c_{1,k}\cdot\sigma^4 + c_{2,k}\cdot\sigma^6~,
\label{eq:fit}
\end{flalign}
where $c_{1,k}$ and $c_{2,k}$ are free fit parameters which depend on the smearing kernel while $R\equiv R_{ud}/|V_{ud}|^{2}$ is the common $\sigma=0$ extrapolation. The quality of the fits is excellent on both volumes and for both regularizations. In the case of the TM regularization, the results of these extrapolations are shown in Figure.~\ref{fig:extrapolation}, again for the two volumes. 

In Table.~\ref{Tab:partial_results}, we report our final determination, for the two considered volumes and for both the TM and OS regularization. For comparison, we also reported in the table the result for $R_{ud}\slash\abs{V_{ud}}^2$ obtained by taking $R_{ud}$ form Ref.~\cite{HFLAV:2022pwe} (HFLAV) and $V_{ud}$ from Ref.~\cite{Hardy:2020qwl} (HT). Although our OS results on the B96 ensemble are still affected by a quite large uncertainty, the spread between the TM and OS results on the B64 ensemble, where the accuracy is less than 2\%, gives a first encouraging indication about the size of the cut-off effects. We are currently performing a more detailed analysis of all systematic effects and plan to extend the analysis to all the physical point ETMC ensembles in order to carry out a reliable continuum limit extrapolation.

\begin{table}
\centering
\begin{tabular}{c c c c} 
  \toprule
    $L$ &  TM ($a=0.08$~fm) & OS ($a=0.08$~fm) & HFLAV+HT ($a=0$)\\
 \midrule
    $5.1$~fm & $3.675(72)$ & $3.550(60)$  & \\     
    $7.6$~fm & $3.562(57)$ & $3.676(236)$ & \\
    $\infty$ &  &  & $3.6615(78)$\\
  \bottomrule
 \end{tabular}
\caption{Preliminary results for $R_{ud}\slash\abs{V_{ud}}^2$ obtained in this work at fixed lattice spacing $a\simeq 0.08$~fm in both the TM and OS lattice regularizations on the volumes $L\simeq 5.1$~fm (ensemble B64) and $L\simeq 7.6$~fm (ensemble B96). For comparison, we also show the result obtained by taking $R_{ud}$ form Ref.~\cite{HFLAV:2022pwe} (HFLAV) and $V_{ud}$ from Ref.~\cite{Hardy:2020qwl} (HT). }
\label{Tab:partial_results}
\end{table}

\section{Conclusion and Outlooks}
We illustrated a method that allows to compute on the lattice the inclusive hadronic decay rates of the $\tau$ lepton without the need of perturbative and/or OPE inputs. We also presented preliminary results for the inclusive decay rate in the $ud$-flavour channel. These have been obtained by applying this method on two $N_f=2+1+1$ QCD gauge ensembles produced by the ETMC with physical pion masses, at fixed cutoff $a=0.07957(13)$~fm and with volumes $L=5.09$~fm and $L=7.64$~fm. 

In our method, as originally proposed in Refs.~\cite{Gambino:2020crt,Gambino:2022dvu}, the step-functions that define the physical phase-space integration domain are smoothed and used as smearing kernels in the algorithm of Ref.~\cite{Hansen:2019idp}. Controlling the limit in which the smearing radius goes to zero is a crucial step of the method, to be performed after the necessary infinite-volume extrapolations. Our numerical results provide a rather solid evidence that this limit can be taken with controlled theoretical and numerical errors. 

We postpone to future work a more detailed illustration of the theoretical analysis of the vanishing smearing width limit and a thorough investigation of all systematic uncertainties, including the required continuum extrapolations. We also plan to extend our computation to the more phenomenologically relevant $us$-flavour channel.

\bibliographystyle{JHEP}
{\small
\bibliography{biblio.bib}

\providecommand{\href}[2]{#2}\begingroup\raggedright\begin{thebibliography}{10}

\bibitem{Aoki:2021kgd}
{\scshape Flavour Lattice Averaging Group (FLAG)} collaboration, \emph{{FLAG
  Review 2021}},
  \href{https://doi.org/10.1140/epjc/s10052-022-10536-1}{\emph{Eur. Phys. J. C}
  {\bfseries 82} (2022) 869}
  [\href{https://arxiv.org/abs/2111.09849}{{\ttfamily 2111.09849}}].

\bibitem{Hudspith:2017vew}
R.J.~Hudspith, R.~Lewis, K.~Maltman and J.~Zanotti, \emph{{A resolution of the
  inclusive flavor-breaking $\tau$ $|V_{us}|$ puzzle}},
  \href{https://doi.org/10.1016/j.physletb.2018.03.074}{\emph{Phys. Lett. B}
  {\bfseries 781} (2018) 206}
  [\href{https://arxiv.org/abs/1702.01767}{{\ttfamily 1702.01767}}].

\bibitem{Maltman:2019xeh}
K.~Maltman et~al., \emph{{Current Status of inclusive hadronic $\tau$
  determinations of |V${_us}$|}},
  \href{https://doi.org/10.21468/SciPostPhysProc.1.006}{\emph{SciPost Phys.
  Proc.} {\bfseries 1} (2019) 006}.

\bibitem{RBC:2018uyk}
{\scshape RBC, UKQCD} collaboration, \emph{{Novel |Vus| Determination Using
  Inclusive Strange \ensuremath{\tau} Decay and Lattice Hadronic Vacuum
  Polarization Functions}}, {\emph{Phys. Rev. Lett.} {\bfseries 121} (2018) }
  [\href{https://arxiv.org/abs/1803.07228}{{\ttfamily 1803.07228}}].

\bibitem{Hansen:2019idp}
M.~Hansen, A.~Lupo and N.~Tantalo, \emph{{Extraction of spectral densities from
  lattice correlators}},
  \href{https://doi.org/10.1103/PhysRevD.99.094508}{\emph{Phys. Rev. D}
  {\bfseries 99} (2019) 094508}
  [\href{https://arxiv.org/abs/1903.06476}{{\ttfamily 1903.06476}}].

\bibitem{Gambino:2020crt}
P.~Gambino and S.~Hashimoto, \emph{{Inclusive Semileptonic Decays from Lattice
  QCD}}, \href{https://doi.org/10.1103/PhysRevLett.125.032001}{\emph{Phys. Rev.
  Lett.} {\bfseries 125} (2020) 032001}
  [\href{https://arxiv.org/abs/2005.13730}{{\ttfamily 2005.13730}}].

\bibitem{Gambino:2022dvu}
P.~Gambino, S.~Hashimoto, S.~M\"achler, M.~Panero, F.~Sanfilippo, S.~Simula
  et~al., \emph{{Lattice QCD study of inclusive semileptonic decays of heavy
  mesons}}, {\emph{JHEP} {\bfseries 07} (2022) }
  [\href{https://arxiv.org/abs/2203.11762}{{\ttfamily 2203.11762}}].

\bibitem{Alexandrou:2022amy}
C.~Alexandrou et~al., \emph{{Lattice calculation of the short and intermediate
  time-distance hadronic vacuum polarization contributions to the muon magnetic
  moment using twisted-mass fermions}},
  \href{https://arxiv.org/abs/2206.15084}{{\ttfamily 2206.15084}}.

\bibitem{Erler:2002mv}
J.~Erler, \emph{{Electroweak radiative corrections to semileptonic tau
  decays}}, {\emph{Rev. Mex. Fis.} {\bfseries 50} (2004) 200}
  [\href{https://arxiv.org/abs/hep-ph/0211345}{{\ttfamily hep-ph/0211345}}].

\bibitem{Bulava:2021fre}
J.~Bulava, M.T.~Hansen, M.W.~Hansen, A.~Patella and N.~Tantalo,
  \emph{{Inclusive rates from smeared spectral densities in the two-dimensional
  O(3) non-linear $\sigma$-model}},
  \href{https://arxiv.org/abs/2111.12774}{{\ttfamily 2111.12774}}.

\bibitem{Alexandrou:2022tyn}
C.~Alexandrou et~al., \emph{{Probing the R-ratio on the lattice}},
  \href{https://arxiv.org/abs/2212.08467}{{\ttfamily 2212.08467}}.

\bibitem{Frezzotti:2004wz}
R.~Frezzotti and G.C.~Rossi, \emph{{Chirally improving Wilson fermions. II.
  Four-quark operators}},
  \href{https://doi.org/10.1088/1126-6708/2004/10/070}{\emph{JHEP} {\bfseries
  10} (2004) 070} [\href{https://arxiv.org/abs/hep-lat/0407002}{{\ttfamily
  hep-lat/0407002}}].

\bibitem{HFLAV:2022pwe}
{\scshape HFLAV} collaboration, \emph{{Averages of $b$-hadron, $c$-hadron, and
  $\tau$-lepton properties as of 2021}},
  \href{https://arxiv.org/abs/2206.07501}{{\ttfamily 2206.07501}}.

\bibitem{Hardy:2020qwl}
J.C.~Hardy and I.S.~Towner, \emph{{Superallowed $0^+ \to 0^+$ nuclear $\beta$
  decays: 2020 critical survey, with implications for V$_{ud}$ and CKM
  unitarity}}, \href{https://doi.org/10.1103/PhysRevC.102.045501}{\emph{Phys.
  Rev. C} {\bfseries 102} (2020) 045501}.

\end{thebibliography}\endgroup
}

\end{document}